\documentclass{elsart4-1}

\usepackage{graphicx}

\usepackage{amssymb,amsmath,amsbsy,mathbbol}

\usepackage[english,francais]{babel}

\newtheorem{e-proposition}[theorem]{Proposition}

\newtheorem{e-definition}[theorem]{Definition\rm}

\renewcommand{\theequation}{\arabic{equation}}
\def\og{\leavevmode\raise.3ex\hbox{$\scriptscriptstyle\langle\!\langle$~}}
\def\fg{\leavevmode\raise.3ex\hbox{~$\!\scriptscriptstyle\,\rangle\!\rangle$}}

\def\tr{\,{\rm tr}\,}

\def\4N{${\mathcal{N}}=4$}

\begin{document}
\begin{flushright}
UUITP-27/04\\
ITEP-TH-50/04\\
\end{flushright}

\begin{frontmatter}


\selectlanguage{english}
\title{Semiclassical Bethe Ansatz and AdS/CFT\thanksref{lab0}}




\selectlanguage{english}
\author{K. Zarembo\thanksref{lab1}}
\thanks[lab0]{Based
on talks at {\it Strings04}, Paris, 28.06.04--02.07.04; {\it 37th
Symposium Ahrenshoop}, Berlin, 23.08.04--27.08.04 and {\it
Integrable Models and Applications}, Sozopol, 31.08.04--04.09.04.}
\thanks[lab1]{Also at ITEP, 117259 Bol. Cheremushkinskaya 25, Moscow, Russia}
\ead{Konstantin.Zarembo@teorfys.uu.se}

\address{Institutionen f\"or Teoretisk Fysik,
Uppsala Universitet\\
Box 803, SE-751 08 Uppsala, Sweden}

\begin{abstract}
The Bethe ansatz can be used to compute anomalous dimensions in \4N
SYM theory. The classical solutions of the sigma-model on
$AdS_5\times S^5$ can also be parameterized by an integral equation
of Bethe type. In this note the relationship between the two Bethe
ans\"atze is reviewed following hep-th/0402207.




\end{abstract}
\end{frontmatter}


\def\N{$\mathcal {N}=4$}
\def\ads{$AdS_5\times S^5$}
\def\pint{{-\!\!\!\!\!\!\int}}
\def\d{\partial}

\selectlanguage{english}
\section{Introduction}

The string dual of the large-$N$ \N\ super-Yang-Mills (SYM) theory
has a geometric description in terms of the \ads\ background with RR
flux \cite{Maldacena:1998re,Gubser:1998bc,Witten:1998qj}. The
duality becomes especially simple at strong coupling or in the
semiclassical limit \cite{Berenstein:2002jq,Gubser:2002tv}. One of
the surprising features of the semiclassical AdS/CFT correspondence
is the appearance of integrable structures on both sides of the
duality. The integrability arises as a quantum symmetry of operator
mixing in CFT \cite{Minahan:2002ve,Beisert:2003tq,Beisert:2003yb}
and as a classical symmetry on the string world-sheet in AdS
\cite{Mandal:2002fs,Bena:2003wd}. This symmetry can be important in
the quantum regime of AdS/CFT \cite{Dolan:2003uh} and seemingly
arises in other examples of the gauge/string theory duality
\cite{Polyakov:2004br}.

The integrability implies that action-angle variables are globally
defined and thus imposes strong restrictions on the classical
dynamics. In many cases the separation of variables can be carried
out even in quantum theory and then the spectrum can be found by
purely algebraic means \cite{FaddeevTakhtajanSklyanin}. Typically,
the spectrum is a Fock space of some elementary excitations whose
creation-annihilation operators satisfy certain quadratic algebra.
The algebra implies that momenta of elementary excitations in a
physical state are subject to a set of algebraic constraints, the
Bethe equations \cite{Bethe:1931hc,Faddeev:1996iy}. The algebraic
Bethe ansatz perhaps is the most generic way to quantize integrable
systems. In the context of AdS/CFT, the Bethe ansatz proved
instrumental in computing perturbative anomalous dimensions
\cite{Beisert:2003xu,Beisert:2003ea,Engquist:2003rn,Serban:2004jf,Kristjansen:2004ei,Kazakov:2004qf,Beisert:2004hm,Lubcke:2004dg,Smedback:1998yn,Freyhult:2004iq,Minahan:2004ds,Callan:2004dt,Kazakov:2004nh,Beisert:2004ag}
and OPE coefficients \cite{RoibanVolovich} of local operators in \N\
SYM. The purpose of these notes, which are based on
\cite{Kazakov:2004qf}, is to explain how classical Bethe equations
arise in string theory on \ads. The quantum counterpart of these
equations, if exists, should describe the exact spectrum of the AdS
string or equivalently the non-perturbative spectrum of large-$N$
SYM. It is not clear  how to derive such quantum Bethe ansatz, but
the success of the discretized string Bethe equations
\cite{Arutyunov:2004vx} in reproducing the near-BMN spectrum of the
string \cite{Callan:2003xr}  can be taken as an indication that
algebraic Bethe ansatz is indeed the right framework to deal with
quantum string theory in \ads.

\section{Integrability in CFT}

I will first explain how Bethe ansatz arises in perturbative SYM. It
is especially useful in the semiclassical limit, which is accurate
for states with large quantum numbers. There are two basic types of
local operators with large quantum numbers in SYM. The majority of
operators have large scaling dimensions at strong coupling
\cite{Gubser:1998bc}, which is the stringy regime of AdS/CFT. This
regime is non-perturbative by definition and hard to access by
conventional field-theory methods. On the other hand, operators with
a large number of constituent fields \cite{Berenstein:2002jq} can
have huge global charges independently of the strength of
interaction and are thus expected to behave stringy even in
perturbation theory. The stringy behavior of both types of operators
 can be qualitatively explained by examining planar diagrams that
dominate their correlation functions. Typical diagrams at strong
coupling diagrams have large number of vertices and propagators and
obviously resemble continuous string world-sheets. Planar diagrams
for large operators always contain many of propagators and resemble
continuous strings even at the lowest orders of perturbation theory.

The field content of \N\ SYM theory consists of gauge fields
$A_\mu$, six scalars $\Phi_i$ and four Majorana fermions
$\Psi_{\alpha}^A$, all in the adjoint representation of $U(N)$. The
action is
\begin{equation}
S=\frac{1}{g^2}\int d^4x\,\tr\left\{
-\frac{1}{2}\,F_{\mu\nu}^2+(D_\mu\Phi_i)^2+[\Phi_i,\Phi_j]^2+{\rm
fermions}\right\}.
\end{equation}
The simplest local gauge-invariant operators are composed of two
types of complex scalar fields $Z=\Phi_1+i\Phi_2$ and
$W=\Phi_3+i\Phi_4$:
\begin{equation}\label{ops}
\mathcal{O}=\tr\left(Z^{L-M}W^M+{\rm permutations}\right).
\end{equation}
 These operators transform non-trivially under an $SU(2)\times U(1)$
subgroup of the $SO(6)$ R-symmetry. $(Z,W)$ transforms as a doublet
of $SU(2)$ and has the $U(1)$ charge $1$. Correlation functions of
operators (\ref{ops}) contain UV divergences and need to be
regularized and renormalized by adding counterterms.  In general the
renormalized operator is a linear combination of several bare
operators: $\mathcal{O}_{\boldsymbol{A}}=
\boldsymbol{Z}_{\boldsymbol{A}}^{\hphantom{\boldsymbol{A}}
\boldsymbol{B}} \mathcal{O}_{\boldsymbol{B}}^{\rm bare}$, where
$\boldsymbol{A}$ and $\boldsymbol{B}$ are multi-indices that
parameterize all possible operators with the same quantum numbers
(the same number of $Z$ and $W$ fields in the present case). The
mixing matrix is defined as
$\Gamma=\boldsymbol{Z}^{-1}d\boldsymbol{Z}/d\ln\Lambda$, where
$\Lambda$ is a UV cutoff.
Its eigenvectors are conformal operators and its eigenvalues are
their anomalous dimensions:
$\Gamma_{\boldsymbol{A}}^{\hphantom{\boldsymbol{A}}\boldsymbol{B}}
\mathcal{O}^{(n)}_{\boldsymbol{B}}=
\gamma_n\mathcal{O}^{(n)}_{\boldsymbol{A}}$, so that the scaling
dimension of the operator $\mathcal{O}^{(n)}$ is
$\Delta_n=L+\gamma_n$. The set of operators (\ref{ops}) is closed
under renormalization. Operators from this set do not mix with
operators that contain $F_{\mu\nu}$, fermions or derivatives
\cite{Beisert:2003jj,Beisert:2004ry}. Including such operators is
possible
\cite{Beisert:2003jj,Beisert:2003yb,Beisert:2003ys,Beisert:2004ry}
but will not be discussed here for the sake of simplicity.

The number of operators of the same length $L$ grows very quickly
with $L$\footnote{Then the number of independent operators with the
same length is exponentially large at $N=\infty$. At finite $N$ the
number of degenerate operators is proportional to some power of
$L$.}, which makes perturbation theory for large operators highly
degenerate. Thus computation of anomalous dimensions is  a
non-trivial problem for large operators even at one loop. The
following parametrization of operators (\ref{ops}) enormously
simplifies this problem. Let us associate the field $Z$ with spin up
and the field $W$ with spin down. An operator of the form
(\ref{ops}) then defines a distribution of spins on a periodic
one-dimensional lattice of length $L$:
$$
\tr ZZZWWZZZWWWZWZZZZ\ldots \qquad\longleftrightarrow\qquad
\left|\uparrow\uparrow\uparrow\downarrow\downarrow
\uparrow\uparrow\uparrow\downarrow\downarrow\downarrow
\uparrow\downarrow\uparrow\uparrow\uparrow\uparrow \ldots
\right\rangle.
$$
The map between the operators and the states of the spin chain is
one-to-one if the states are required to be translationally
invariant. The mixing matrix acts linearly on the operators and thus
can be interpreted as a Hamiltonian of a spin chain.

\begin{figure}[t]
\centerline{\includegraphics[width=8cm]{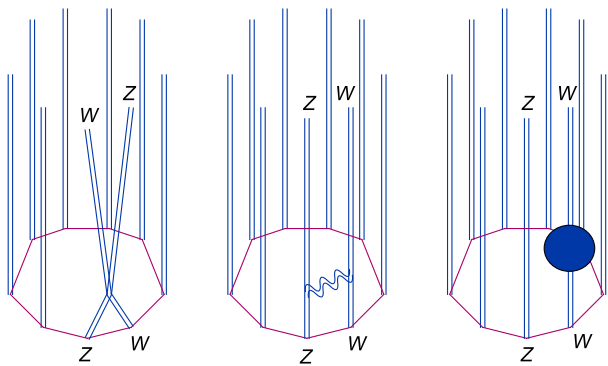}}
\caption{\label{mm}\small The one-loop diagrams.}
\end{figure}

The one-loop mixing matrix can be easily computed. The three
diagrams that contribute at this order are shown in fig.~\ref{mm}.
The gluon exchange and the self-energy produce the same renormalized
operator, while the scalar vertex can lead to the interchange of $Z$
and $W$ fields. At large $N$ the interchange can only occur between
adjacent sites of the lattice. Indeed, an insertion of the vertex
between a pair propagators produces a non-planar graph unless the
propagators start from the adjacent sites. The planar mixing matrix
is thus a Hamiltonian of a spin chain with nearest-neighbor
interactions. Explicitly \cite{Minahan:2002ve},
\begin{equation}\label{}
\Gamma=\frac{\lambda}{8\pi^2}\sum_{l=1}^{L} (1-P_{l,l+1}),
\end{equation}
where $\lambda=g^2N$ is the 't~Hooft coupling and $P$ is the
permutation operator: $P\,a\otimes b=b\otimes a$. The use of the
identity $P=(1+\boldsymbol{\sigma}\otimes\boldsymbol{\sigma})/2$
brings the mixing matrix to the familiar form of the Heisenberg
Hamiltonian:
\begin{equation}\label{}
\Gamma=\frac{\lambda}{16\pi^2}\sum_{l=1}^{L}
(1-\boldsymbol{\sigma}_l\cdot\boldsymbol{\sigma}_{l+1}).
\end{equation}
The physical states that describe operators in the SYM theory should
satisfy an additional constraint:
\begin{equation}\label{}
 U\left| {\rm phys}\right\rangle=\left| {\rm phys}\right\rangle,
\end{equation}
where $U$ is the shift operator:
$U^{-1}\boldsymbol{\sigma}_lU=\boldsymbol{\sigma}_{l-1}$. The
eigenvalues of $U$ define the total momentum which must be zero (or
an integer multiple of $2\pi$) for translationally invariant
physical states.

\begin{figure}[t]
\centerline{\includegraphics[width=15cm]{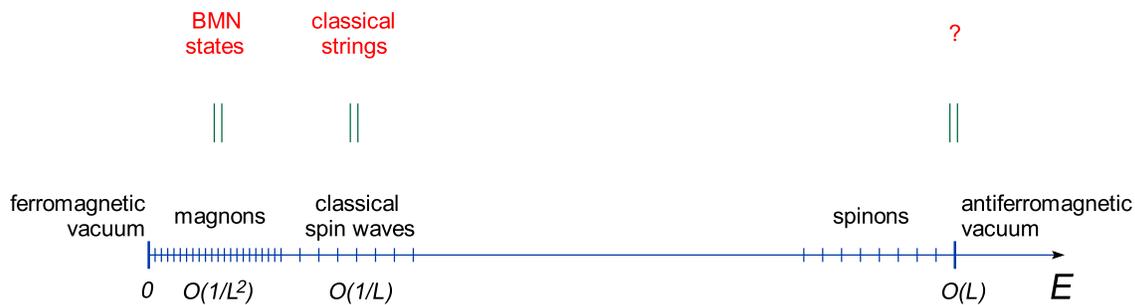}}
\caption{\label{spec}\small The spectrum of the Heisenberg spin
chain.}
\end{figure}

Though Heisenberg Hamiltonian contains no adjustable parameters
except for the length of the chain it is possible to identify
several energy scales in its spectrum (fig.~\ref{spec}). The ground
state is the ferromagnetic vacuum (all spins up) and corresponds to
a chiral primary operator  $\tr Z^L$. This operator belongs to a
short multiplet of $\mathcal{N}=4$ supersymmetry and has zero
anomalous dimension to all orders in perturbation theory. In
particular it has zero anomalous dimension at one loop, so the
ground state of the spin chain has zero energy. The excited states
are obtained by flipping one or more spins. In some approximation
the spectrum is generated by
\begin{equation}\label{add}
a^\dagger_n=\frac{1}{\sqrt{L}}\sum_{l=1}^L \e^{\frac{2\pi
inl}{L}}\sigma^-_l.
\end{equation}
The operator $a^\dagger_n$ creates a magnon with the mode number $n$
and the momentum $p=2\pi n/L$. If $L$ is sufficiently large, Fock
states $a^\dagger_{n_1}\ldots a^\dagger_{n_M}\left|0\right\rangle$
(with $M\ll L$) approximate the eigenstates of the Heisenberg
Hamiltonian up to computable $1/L$ corrections \cite{Callan:2004dt}.
The multi-magnon states correspond to the BMN operators
\cite{Berenstein:2002jq} and are dual to string states in the
pp-wave limit of the \ads\ geometry \cite{Metsaev}. The anomalous
dimensions of the BMN operators (the energies of magnons),
\begin{equation}\label{bmn}
\gamma=\frac{\lambda}{2L^2}\sum_{k=1}^M n_k^2,
\end{equation}
match with the energies of the string oscillators
\cite{Berenstein:2002jq}. The zero-momentum condition becomes the
level matching condition on the string side.

The situation changes when the number of magnons becomes
macroscopically large: $M\sim L$. Then the interaction between
magnons cannot be neglected any more and the Fock space generated by
simple operators (\ref{add}) is no longer a good approximation for
the spectrum. The remarkable property of the Heisenberg model, which
stems from its completely integrability, is that the exact spectrum
is still a Fock space. The simple creation operators (\ref{add}) get
dressed by interactions, but the exact spectrum-generating operators
still obey relatively simple exchange relations (the Yang-Baxter
algebra) and the spectrum can be computed algebraically.
The energy eigenstates as
before are parameterized by momenta of individual magnons, but now
the momenta satisfy a set of algebraic equations
\cite{Bethe:1931hc,Faddeev:1996iy}:
\begin{equation}\label{bethe}
\left(u_j+i/2\over u_j-i/2\right)^L= \prod_{k\ne j}
 {u_j-u_k+i\over
u_j-u_k-i}.
\end{equation}
The Bethe roots $u_j$, $j=1,\ldots ,M$ are distinct complex numbers
which parameterize the momenta of magnons:
$$
\e^{ip}={u+i/2\over u-i/2}.
$$
The momentum constraint takes the form
\begin{equation}\label{}
\prod_{j}{u_j+i/2\over u_j-i/2}=1,
\end{equation}
and the anomalous dimension is computed as
\begin{equation}\label{}
\gamma={\lambda\over 8\pi^2}\sum_{j}{1\over u_j^2+1/4}\,.
\end{equation}

\begin{figure}[t]
\centerline{\includegraphics[width=10cm]{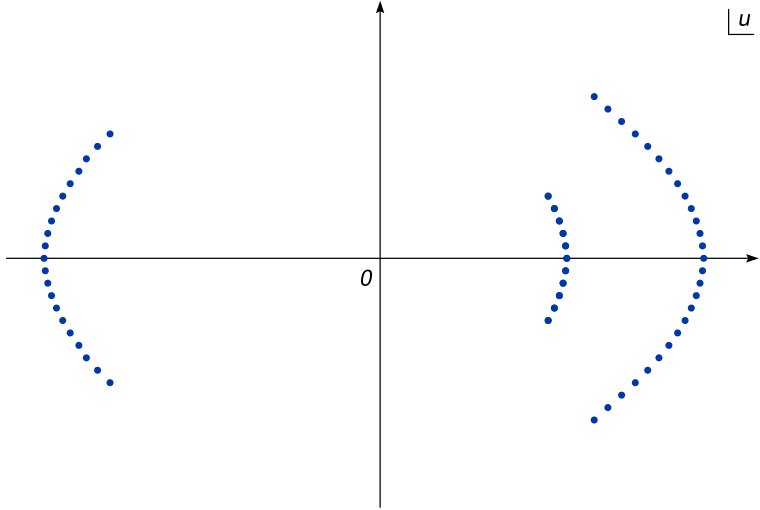}}
\caption{\label{roots}\small Macroscopically large Bethe strings.}
\end{figure}

The simplest solution of the Bethe equations that satisfies the
momentum constraint contains two roots, which describes two magnons
with opposite momenta:
\begin{equation}\label{}
u_1=-u_2=\frac{1}{2}\,\cot\frac{\pi n}{L-1}, \qquad n=1,2,\ldots
\end{equation}
The anomalous dimension is
\begin{equation}\label{}
\gamma=\frac{\lambda}{\pi^2}\,\sin^2\frac{\pi n}{L-1}\,.
\end{equation}
When $L$ is large, $u_{1,2}\approx \pm L/2\pi n$ and we get back to
the BMN formula (\ref{bmn}). The fact that Bethe roots scale as
$u_i\sim L$ is true for all BMN-like states. The right hand side of
Bethe equations can then be replaced by $1$, or in other words the
interaction between magnons can be neglected.
A little complication occurs when more than one magnon occupies the
same momentum state. Then the above argument does not apply since
magnons cannot have the same rapidities.
As a result, the rapidities split in the complex plane and magnons
with the same momentum form a bound state. Because $(u+i/2)/(u-i/2)$
is no longer a pure phase, the left hand side of (\ref{bethe}) turns
to zero or to infinity as $L\rightarrow\infty$. This should be
compensated by a zero or a pole on the right hand side, which can
only happen if two or more rapidities are separated by $\pm i$. The
rapidities of magnons in the bound state thus form a rigid array
with two or more roots at $u_{\rm c.m.} +ir$, where $r$ are integers
or half-integers. Such arrays are usually called strings. The number
of roots in a string can be arbitrary, even macroscopically large
\cite{sutherland} (fig.~\ref{roots}). In the latter case strings
bend on the macroscopic scales and form some contours in the complex
plane. The corresponding Bethe states describe macroscopic spin
waves and are dual to semiclassical strings in \ads\
\cite{Frolov:2003qc,Beisert:2003xu}.

Since Bethe roots scale linearly with $L$, it is natural to define
$x_i=u_i/L$, which stays finite at $L\rightarrow\infty$. Taking the
logarithm of (\ref{bethe}) and expanding in $1/L$, we get
\begin{equation}\label{}
{1\over x_j}+2\pi n_j={1\over L}\sum_{k\ne j} {2\over x_j-x_k},
\end{equation}
where the phases $2\pi n_j$ parameterize the branches of the
logarithm. The mode numbers $n_j$ are different for different Bethe
strings.
The distance between the adjacent roots scales as $x_{k}-x_{k+1}\sim
1/L$, so the distribution of roots can be characterized by a
continuous density in the scaling limit:
\begin{equation}\label{}
\rho(x)=\frac{1}{L}\sum_j\delta(x-x_j).
\end{equation}
The density is defined on a collection of contours $C=C_1\cup\ldots
\cup C_K$ in the complex plane and is normalized as
\begin{equation}\label{nom}
\int_C dx\rho(x)=\frac{M}{L}\,,
\end{equation}
where $M$ is the total number of magnons, or the number of $W$
fields in the operator (\ref{ops}). Equivalently, the distribution
of Bethe roots can be characterized by the resolvent:
\begin{equation}\label{g}
G(x)=\frac{1}{L}\sum_j\frac{1}{x-x_j}=\int_C
dy\,\frac{\rho(y)}{x-y}\,.
\end{equation}
The Taylor expansion of $G(x)$ at zero generates local conserved
charges of the Heisenberg model
\cite{Arutyunov:2003rg,Engquist:2003rn}. In particular, the total
momentum is $-G(0)$. Translational invariance then requires
\begin{equation}\label{mom}
-G(0)=\int_C dx\,\frac{\rho(x)}{x}=2\pi m.
\end{equation}
The next Taylor coefficient determines the anomalous dimension:
\begin{equation}\label{adim}
\Delta -L=\frac{\lambda}{8\pi^2L}\int_C dx\,\frac{\rho(x)}{x^2}\,.
\end{equation}
The Bethe equations reduce to a singular integral equation for the
density:
\begin{equation}\label{bte}
\pint_{{\bf C}} {dy\,\rho(y)\over x-y} = {1\over x}+2\pi n_k, \qquad
x\in C_k,
\end{equation}
which resembles the saddle-point equation for a distribution of
eigenvalues of a random matrix \cite{Brezin:1977sv}. Its general
solution is known and can be expressed in terms of the hyperelliptic
integrals \cite{Kazakov:2004qf}. The associated Riemann surface
is obtained by gluing
together two copies of the complex plane with cuts along the
contours $C_k$.

The one- and two-cut solutions (rational and elliptic cases) were
worked out in \cite{Beisert:2003xu,Beisert:2003ea,Kazakov:2004qf}
and were compared to the Frolov-Tseytlin string solutions
\cite{Frolov:2003qc}. $1/L$ corrections, that can be explicitly
calculated in the simplest rational case \cite{Lubcke:2004dg}, on
the string side correspond to quantum corrections on the world sheet
\cite{Frolov:2003tu}. The scaling dimensions of operators were found
to agree with the energies of the string states up to two loops. At
three loops the agreement breaks down for the BMN states
\cite{Callan:2003xr} and for the macroscopic strings
\cite{Serban:2004jf,Beisert:2004hm,BeisertStrings},
but the two-loop agreement can be established quite generally at the
level of the effective actions \cite{Kruczenski:2003gt} or the
equations of motion \cite{Mikhailov:2003gq}. Even the higher charges
of the integrable hierarchies, that do not have geometric
interpretation in AdS/CFT, were found to agree
\cite{Arutyunov:2003rg,Engquist:2004bx}. The classical solutions of
the sigma-model can also be parameterized by an integral equation of
Bethe type. Such an equation was derived for strings on $S^3\times
\mathbb{R}^1$ \cite{Kazakov:2004qf}, $AdS_3\times S^1$
\cite{Kazakov:2004nh} and $S^5\times \mathbb{R}^1$
\cite{Beisert:2004ag}. I will discuss the first case.

 \section{Integrability in AdS}
\label{adssec}

The R-charges of the operator $\tr(Z^{L-M}W^M+\ldots )$, $L$ and
$M$, are dual to angular momenta of the string on $S^5$. The string
in the middle of $AdS_5$ has two non-zero angular momenta if it
moves in $S^3\subset S^5$. The world sheet is then parameterized by
the global AdS time $X^0$ and by four Cartesian coordinates $X^i$
constrained by $X^iX^i=1$. A point on the three-sphere defines a
group element of $SU(2)$:
\begin{equation}\label{su2}
g = \left(
\begin{array}{cc}
  X_1+iX_2 & X_3+iX_4 \\
 -X_3+iX_4 & X_1-iX_2
\end{array}
\right) \equiv \left(
\begin{array}{cc}
  Z_1 & Z_2 \\
  -{\bar Z}_2 & {\bar Z}_1
\end{array}\right) \in SU(2),
\end{equation}
and the equations of motion of the string can be conveniently
formulated in terms of the currents
\begin{equation}\label{cur}
j_a=g^{-1}\d_a g={\sigma^A\over 2i}\,j^A_a.
\end{equation}
The relevant part of the string action takes the following form in
the conformal gauge\footnote{The world-sheet metric is $(+-)$.}:
\begin{equation}\label{sm}
S_{\sigma m}= -\frac{\sqrt{\lambda}}{4\pi} \int_{0}^{2\pi} d\sigma
\int d\tau\, \left[\frac{1}{2}\,{\rm Tr} j_a^2+(\d_a X_0)^2\right].
\end{equation}
The equations of motion are
\begin{equation}\label{forX}
\d_+\d_- X_0=0,
\end{equation}
\begin{equation}
 \label{chi1} \d_+j_-+\d_-j_+=0,
\end{equation}
where $\partial_{\pm}=\partial_\tau\pm\partial_\sigma$ and
$j_\pm=j_\tau\pm j_\sigma$. The currents are flat as a consequence
of their definition:
\begin{equation}
\label{chi2} \d_+j_--\d_-j_++[j_+,j_-]=0,
\end{equation}
 The equations of motion should be
supplemented by Virasoro constraints
\begin{equation}\label{vir}
\frac{1}{2}\,\tr j_\pm^2=-(\partial_\pm X^0)^2.
\end{equation}
Since (\ref{forX}) is trivially solved by
$$
X^0=\kappa\tau,
$$
we find:
\begin{equation}\label{}
\frac{1}{2}\,\tr j_\pm^2=-\kappa^2.
\end{equation}

 The global symmetry of the sigma-model (\ref{sm}) is
$SU_L(2)\times SU_R(2)\times\mathbb{R}$. The first two factors are
associated with the left and right group multiplications:
$g\rightarrow hg$ and $g\rightarrow gh$. The Noether currents of
these symmetries are $l_a$ and $j_a$, where $j_a$ is defined in
(\ref{cur}) and
\begin{equation}\label{}
l_a=gj_ag^{-1}=\d_a g\, g^{-1}={\sigma^A\over 2i}\,l^A_a.
\end{equation}
Therefore the Noether charges
\begin{equation}\label{}
Q_L^A=\frac{\sqrt{\lambda}}{4\pi}\int_0^{2\pi} d\sigma\, l^A_\tau,
\qquad Q_R^A=\frac{\sqrt{\lambda}}{4\pi}\int_0^{2\pi} d\sigma\,
j^A_\tau
\end{equation}
generate the left and right group shifts. The dual R-charges in the
SYM theory can be easily identified. The scalars of the SYM and the
Cartesian coordinates on the sphere transform in the same way under
$SO(6)$: $\Phi_i\sim X^i$. Since $Z=\Phi_1+i\Phi_2$ and
$W=\Phi_3+i\Phi_4$, these fields transform as $Z_1$ and $Z_2$ in
(\ref{su2}). Thus $(Z_1,Z_2)$ and $(Z,W)$ are doublets of $SU_R(2)$,
so that $Z$ has $Q^3_R=1$ and $W$ and $Q^3_R=-1$. For the string
dual of the operator (\ref{ops}) we thus have
\begin{equation}\label{}
Q^3_R=L-2M.
\end{equation}
Under the left shifts $(Z_1,-\bar{Z}_2)$ and $(Z_2,-\bar{Z}_1)$
transform as doublets. Therefore, $(Z,-\bar{W})$ and $(W,-\bar{Z})$
are doublets of $SU_L(2)$ and both fields $Z$ and $W$ have
$Q^3_L=1$. Hence the left charge of the operator (\ref{ops}) is just
the length of the spin chain:
\begin{equation}\label{}
Q^3_L=L.
\end{equation}
The time translations $X^0\rightarrow X^0+t$ generate scale
transformations on the boundary of $AdS_5$, and the energy of the
string should be identified with the scaling dimension of the dual
operator:
\begin{equation}\label{}
\Delta=\frac{\sqrt{\lambda}}{2\pi}\int_0^{2\pi} d\sigma\, \d_\tau
X_0 =\sqrt{\lambda}\,\kappa.
\end{equation}

The equations of motion for the chiral field (\ref{chi1}),
(\ref{chi2}) are completely integrable \cite{Pohlmeyer:1975nb} and
can be effectively linearized with the help of the inverse
scattering transformation \cite{book_of_soliton}. The method is
based on the zero-curvature representation \cite{Faddeev's_book}
introduced for the sigma-model in \cite{Zakharov:pp}. The rescaled
currents \cite{Zakharov:pp}
\begin{equation}\label{cn}
J_\pm(x) = \frac{j_\pm}{1\mp x}
\end{equation}
are flat for any value of  $x$:
\begin{equation}\label{flat}
\d_+J_--\d_-J_++[J_+,J_-]=0,
\end{equation}
as a consequence of the equations of motion. On the other hand, if
(\ref{flat}) is satisfied for any $x$, then $j_\pm$ are solutions of
the equations of motion. The equations of motion (\ref{chi1}) and
(\ref{chi2}) and the flatness condition (\ref{flat}) are thus
completely equivalent.

The monodromy of the flat connection (\ref{cn}) defines the
quasi-momentum $p(x)$:
\begin{equation}\label{mmat}
\Omega(x)=P\exp\left(-\int_0^{2\pi} d\sigma\,J_\sigma\right)
=P\exp\int_0^{2\pi} d\sigma\, \frac{1}{2}\left( {j_+\over x-1} +
{j_-\over x+1}\right),
\end{equation}
\begin{equation}\label{defp}
\tr\Omega(x)=2\cos p(x),
\end{equation}
where the integral is taken along a fixed-time section of the
world-sheet. The quasi-momentum is a functional of the world-sheet
currents and potentially depends on time. The key point is that it
becomes time-independent on-shell. Indeed, the trace of the holonomy
of a flat connection does not depend on the contour of integration.
Shifting the time slice along which the flat connection is
integrated changes nothing and therefore the quasi-momentum is
conserved as soon as the equations of motion are satisfied. It can
thus be regarded as a generating function for an infinite set of
integrals of motion.

The $SU_{L,R}(2)$ charges appear in the expansion of $p(x)$ at zero
and at infinity. At infinity: $J_\sigma(x)=j_\tau/x+\ldots $, so
\begin{equation}\label{}
\tr\Omega=2+\frac{1}{2x^2}\int_0^{2\pi}d\sigma_1 d\sigma_2\, \tr
j_\tau (\sigma_1) j_\tau (\sigma_2)+\ldots =2-\frac{4\pi^2
Q_R^2}{\lambda x^2}+\ldots =2-\frac{4\pi^2 (L-2M)^2}{\lambda
x^2}+\ldots \,.
\end{equation}
Hence,
\begin{equation}\label{pinfty}
p(x)=-\frac{2\pi(L-2M)}{\sqrt{\lambda}\, x}+\ldots\qquad
(x\rightarrow\infty).
\end{equation}
Further terms in the expansion of the monodromy matrix $\Omega(x)$
generate Yangian charges, which potentially play an important role
in the AdS/CFT correspondence \cite{Dolan:2003uh,DolanNappi}.

At $x\rightarrow 0$, $\partial_\sigma+J_\sigma(x)=\partial_\sigma
+j_\sigma-xj_\tau+\ldots=g^{-1}(\partial_\sigma -xl_\tau+\ldots )g$,
so
\begin{equation}\label{}
\tr\Omega=2+\frac{x^2}{2}\int_0^{2\pi}d\sigma_1 d\sigma_2\, \tr
l_\tau (\sigma_1) l_\tau
(\sigma_2)+\ldots=2-\frac{4\pi^2Q_L^2}{\lambda }\,x^2
+\ldots=2-\frac{4\pi^2L^2}{\lambda }\,x^2 +\ldots\,,
\end{equation}
which yields
\begin{equation}\label{p0}
p(x)=2\pi m+\frac{2\pi L}{\sqrt{\lambda} }\, x+\ldots\qquad
(x\rightarrow 0),
\end{equation}
where $m$ is an arbitrary integer.

The local charges can be obtained by expanding the quasi-momentum in
the Lorant series at $x=\pm 1$ for a systematic recursive procedure
exists \cite{Faddeev's_book}. A slightly different but equivalent
definition of the quasi-momentum is more appropriate for that
purpose. Let us consider the auxiliary linear problem
\begin{equation}\label{dir}
\left[ \d_\sigma - \frac{1}{2} \left({j_+\over x-1} + {j_-\over x+1}
\right) \right] \psi=0,
\end{equation}
where $\psi(\sigma;x)$ is a two-component vector ($j_\pm$ are
anti-Hermitian $2\times 2$ matrices). This equation can be regarded
as a  spectral problem for a one-dimensional Dirac operator with a
periodic potential, where $x$ plays the role of the spectral
parameter. Two linearly independent solutions of this spectral
problem can be chosen quasi-periodic: $\psi(\sigma+2\pi;x)=\e^{\pm
ip(x)}\psi(\sigma;x)$. This is the standard definition of the
quasi-momentum. It is easy to see that the previous definition is
equivalent to that. Indeed,
$\psi(\sigma+2\pi;x)=\Omega(x)\psi(\sigma;x)$ in any basis, even in
which the wave function is not quasi-periodic. Requirement of
quasi-periodicity is equivalent to diagonalization of the monodromy
matrix whose eigenvalues are precisely $\e^{\pm ip(x)}$.

When $x$ is close to $+1$ or $-1$, the linear problem (\ref{dir})
can be solved in the WKB approximation, because it takes  the form
\begin{equation}\label{sc}
 \left(i\hbar\partial _\sigma +V\right)\psi =0,
\end{equation}
where $\hbar\equiv 4(x\pm 1)\rightarrow 0$ and
\begin{equation}\label{defM}
 V=u^A\sigma ^A,\qquad u^A=-j^A_\mp\pm\frac{\hbar}{8\mp\hbar}\,j^A_\pm.
\end{equation}
Plugging the WKB ansatz $\psi=\e^{iS/\hbar}\chi $ into (\ref{sc}),
we find that
\begin{equation}\label{se}
\left(V-\partial_\sigma S\right)\chi=0.
\end{equation}
Thus $\chi $ is an eigenvector of $V$ with the eigenvalue $\partial
_\sigma S$. Using the Virasoro constraints (\ref{vir}) we find that
the two eigenvalues of $V$ are $\pm 2\kappa+O(\hbar)$. Choosing the
upper sign we get
 $S(\sigma)=\kappa\sigma$,
 $\psi =\,{\rm e}\,^{i\kappa \sigma /2(x\pm 1)}$ and
$$
\psi(\sigma+2\pi;x)=\exp\left(\frac{i\pi\kappa }{x\pm
1}\right)\psi(\sigma;x).
$$
Therefore\footnote{The local analysis determines $p(x)$ only up to a
sign. Fixing this sign ambiguity singles out a particular class of
solutions. Some solutions (pulsating strings
\cite{Minahan:2002rc,Khan:2003sm,Smedback:1998yn,Engquist:2003rn})
which are not consistent with the particular choice of signs in
(\ref{p1}) are discussed in sec.~5.3 of \cite{Kazakov:2004qf}.}
\begin{equation}\label{p1}
p(x)=-\frac{\pi\kappa}{x\pm 1}+\ldots \qquad (x\rightarrow\mp 1).
\end{equation}

Let me make a short digression on the higher orders of the WKB
expansion. The first non-trivial charges arise at
$O(\hbar^0)$\footnote{This is a special property of the $SU(2)$
sector. In the full \ads\ sigma-model, even the zeroth charge
($O(1/\hbar)$ term in the quasi-momentum) is non-trivial
\cite{Arutyunov:2004yx}.} and can be interpreted as the energy and
momentum \cite{Faddeev:1985qu}. Those are not the standard Noether
charges one would get from the sigma-model action (\ref{sm}) and
they generate the equations of motion (\ref{chi1}), (\ref{chi2})
only if the Poisson structure is properly modified
\cite{Faddeev:1985qu}. Since these facts may have important
implications for quantization, I will rederive them from the
observation that the $O(\hbar^0)$ correction to the quasi-momentum
is the Berry's phase. The $O(\hbar)$ charges are explicitly
calculated in \cite{Beisert:2004ag} for a more general case of the
$SO(6)$ sigma-model.

 The
monodromy matrix, defined as
\begin{equation}\label{}
T(\sigma ;x)= P\exp\int_0^{\sigma } d\tilde\sigma\,
\frac{1}{2}\left( {j_+\over x-1} + {j_-\over x+1}\right),
\end{equation}
is a solution of (\ref{dir}) with the initial condition
$T(0;x)={\mathbb 1}$. This object is a generalization of
(\ref{mmat}): $\Omega(x) =T(2\pi ;x)$. In the semiclassical
approximation,
\begin{equation}\label{semimon}
 T(\sigma)=\sum_{n}\exp\left({\frac{i}{\hbar}\,\int_{0}^{\sigma }
 d\tilde\sigma \,\mu _n(\tilde\sigma )}
 \right)
 \chi _n(\sigma )
 \chi_n ^\dagger(0),
\end{equation}
where $\chi _n(\sigma )$ and $\mu_n (\sigma )$ are the normalized
eigenvectors and eigenvalues of $V(\sigma )$:
$$V\chi _n=\mu _n\chi _n,$$
which depend on $\sigma $ as a parameter. The representation
(\ref{semimon}) follows from (\ref{se}) and is accurate up to
$O(\hbar)$ corrections.

Though matrix elements of $V$ are periodic functions of $\sigma $,
the eigenvectors of $V$ are only quasi-periodic. It is well known
that the transport of an eigenvector $\chi _n$ around a closed
contour in the parameter space generates a geometric Berry's phase
\cite{Berry:1984jv}:
\begin{equation}\label{Bphase}
 \chi _n(\sigma +2\pi )=\e^{i\gamma _n}\chi _n(\sigma ),
\end{equation}
which cannot be written as a local functional of the matrix elements
of $V(\sigma )$. The best one can do is analytically continue
$V(\sigma )$ into the interior of the unit disc whose boundary is
the circle parameterized by $\sigma $. Then \cite{Berry:1984jv}
\begin{equation}\label{Berry}
 \gamma _n=2{\mathop{\mathrm{Im}}\nolimits}
 \int_{0}^{2\pi }d\sigma  \,\int_{0}^{1}d\xi  \,
 \sum_{m\neq n}\frac{\left\langle n\right|
 \frac{\partial V}{\partial \xi }\left| m\right\rangle\,
 \left\langle m\right|\frac{\partial V}{\partial \sigma }
 \left| n\right\rangle}{(\mu _n-\mu _m)^2}\,,
\end{equation}
where $\xi $ is the radial variable. The phase mod $2\pi $ does not
depend on how $V$ is continued into the interior of the disc as long
as the continuation is analytic.

Because $V$ is traceless, its two eigenvalues are equal in magnitude
and opposite in sign: $\mu _1=-\mu _2\equiv \mu $. The same is true
for the Berry's phases: $\gamma _1=-\gamma _2\equiv \gamma $. From
(\ref{defp}), (\ref{semimon}) and (\ref{Bphase}) we find that
\begin{equation}\label{}
 p(x)=-\frac{1}{4(x\pm 1)}\int_{0}^{2\pi }
 d\sigma \,\mu(\sigma ;x)-\gamma(x)+O\left(x\pm 1\right).
\end{equation}
There are two sources of $O(1)$ terms in the quasimomentum: the
Berry's phase and the subleading term in the expansion of the
eigenvalue $\mu =u\equiv\sqrt{u^Au^A}$:
\begin{equation}\label{}
 \mu =2\kappa \mp\frac{\hbar}{16\kappa }\,j_+^Aj_-^A+O(\hbar^2).
\end{equation}
The Berry's phase can be calculated from (\ref{Berry}):
\begin{equation}\label{}
 \gamma =\frac{1}{2}
 \int_{0}^{2\pi }d\sigma \,\int_{0}^{1} d\xi \,
 \varepsilon ^{ABC}\,\frac{u^A}{u^3}\,
 \frac{\partial u^B}{\partial \xi }
 \frac{\partial u^C}{\partial \sigma }
 =-\frac{1}{16\kappa^3 }
 \int_{0}^{2\pi }d\sigma \,\int_{0}^{1} d\xi \,
 \varepsilon ^{ABC}j_\mp^A\,
 \frac{\partial j_\mp^B}{\partial \xi }
 \frac{\partial j_\mp^C}{\partial \sigma }
 +O(\hbar)
\end{equation}
This gives two conserved charges:
\begin{equation}\label{p+-}
 P_\pm=\int_{0}^{2\pi }d\sigma \left(
 \frac{1}{4\kappa ^2}\int_{0}^{1} d\xi \,
 \varepsilon ^{ABC}j_\pm^A\,
 \frac{\partial j_\pm^B}{\partial \xi }
 \frac{\partial j_\pm^C}{\partial \sigma }
 \pm \frac{1}{4}\,j_+^Aj_-^A
 \right).
\end{equation}
Their linear combinations can be interpreted as the energy and
momentum \cite{Faddeev:1985qu}\footnote{The local charges are given
in \cite{Faddeev:1985qu} in the local form which is not manifestly
$SU(2)$ invariant. That representation is completely
 equivalent to (\ref{p+-}) \cite{Faddeev's_book}.}:
\begin{eqnarray}\label{m}
 H&=&\int_{0}^{2\pi }d\sigma \left[\frac{1}{4\kappa ^2}
 \int_{0}^{1} d\xi \,
 \varepsilon ^{ABC}\left(j_+^A\,
 \frac{\partial j_+^B}{\partial \xi }
 \frac{\partial j_+^C}{\partial \sigma }
 -j_-^A\,
 \frac{\partial j_-^B}{\partial \xi }
 \frac{\partial j_-^C}{\partial \sigma }\right)
 +\frac{1}{2}\,j_+^Aj_-^A
 \right],
  \\ \label{e}
 P&=&\frac{1}{4\kappa ^2}
 \int_{0}^{2\pi }d\sigma \int_{0}^{1} d\xi \,
 \varepsilon ^{ABC}\left(j_+^A\,
 \frac{\partial j_+^B}{\partial \xi }
 \frac{\partial j_+^C}{\partial \sigma }
 +j_-^A\,
 \frac{\partial j_-^B}{\partial \xi }
 \frac{\partial j_-^C}{\partial \sigma }\right).
\end{eqnarray}
These are {\it not} the standard momentum and energy of the
sigma-model (\ref{sm}). The latter are trivial as long as the
Virasoro constraints are imposed. $P$ generates translations and $H$
generates the equations of motion (\ref{chi1}), (\ref{chi2}), if the
Poisson brackets of the currents are defined to be
\begin{equation}\label{pois}
 \{j_\pm^A(\sigma ),j_\pm^B(\sigma ')\}=\varepsilon ^{ABC}
 \delta (\sigma -\sigma ')j_\pm^C(\sigma ),
 \qquad
 \{j_+^A(\sigma ),j_-^B(\sigma ')\}=0.
\end{equation}
One would get different Poisson brackets, which contain $\delta
'(\sigma -\sigma ')$, from the action (\ref{sm}). It was argued in
\cite{Faddeev:1985qu} that the non-standard Poisson structure
(\ref{pois}) is more consistent with integrability than the standard
Poisson structure and is better suited for quantization when the
constraints (\ref{vir}) are imposed.

Let us now return to the linear problem (\ref{dir}). Its spectrum
has a band structure, but since the Dirac operator in (\ref{dir})
does not possess any particular Hermiticity properties, the allowed
bands do not lie on the real axis. Instead, they form symmetric
contours in the complex plane. The quasi-momentum is real:
$\tr\Omega\in\mathbb{R}$ and $|\tr\Omega|<2$ on the allowed zones.
Forbidden zones can be defined as contours on which the
quasi-momentum is pure imaginary: $\tr\Omega\in \mathbb{R}$,
$|\tr\Omega|>2$. At zone boundaries $\tr\Omega=2$ and the monodromy
matrix $\Omega$ degenerates into a Jordan cell.
The two quasi-periodic solutions of the Dirac equation, let us
denote them $\psi_\pm(\sigma;x)$, become degenerate there.
Encircling a zone boundary in the complex plane of $x$ interchanges
the two solutions, so zone boundaries are branch points of
$\psi_\pm(x)$. By cutting two copies of the complex plane along the
forbidden zones and gluing them together we get a Riemann surface on
which $\psi_\pm$ are globally defined as two branches of a single
meromorphic function. The same is true for the eigenvalues of the
monodromy matrix $\e^{\pm ip(x)}$. The trace of the monodromy matrix
$\tr\Omega(x)$ has an essential singularity at $x=\pm 1$, where the
Dirac operator has a pole, but otherwise $\tr\Omega(x)$ is a
holomorphic function of $x$. The branch points arise when we solve
$\e^{ip}+\e^{-ip}=\tr\Omega$ for $\e^{ip}$ precisely at the zone
boundaries where $\tr\Omega=2$.

A particular branch of the quasi-momentum $p(x)$ is an analytic
function of $x$ on the complex plane with cuts and has single poles
at $x=\pm 1$. Subtracting the poles we get a function
\begin{equation}\label{defG}
G(x)=p(x)+\frac{\pi\kappa}{x-1}+\frac{\pi\kappa}{x+1},
\end{equation}
which has only branch cut singularities and therefore is completely
determined by its discontinuities: $G(x+i0)-G(x-i0)\equiv
2i\pi\rho(x)$. It is straightforward to prove that $G(x)$ admits the
dispersion representation
\begin{equation}\label{sr}
G(x)=\int_C dy\,\frac{\rho(y)}{x-y}\,.
\end{equation}

As a consequence of the spectral representation (\ref{sr}), the
density $\rho(x)$ satisfies an integral equation which reflects the
unimodularity of the monodromy matrix. To derive this equation, let
us consider the behavior of the quasi-momentum near a forbidden zone
where the quasi-momentum experiences a jump. The two branches of the
double-valued analytic function $\e^{ip(x\pm i0)}$ are the two
eigenvalues of the monodromy matrix $\Omega(x)$ and thus satisfy
$\e^{ip(x+i0)}\e^{ip(x-i0)}=1$, or
\begin{equation}\label{}
p(x+i0)+p(x-i0)=2\pi n_k,\qquad x\in  C_k,
\end{equation}
on all of the forbidden zones. Taking into account (\ref{defG}) and
(\ref{sr}), we get
\begin{equation}\label{sbethe}
2\pint dy\,\frac{\rho(y)}{x-y}=\frac{2\pi\kappa}{x-1}
+\frac{2\pi\kappa}{x+1}+2\pi n_k, \qquad x\in C_k.
\end{equation}
The density also satisfies several normalization conditions which
follow from (\ref{pinfty}), (\ref{p0}) and (\ref{p1}):
\begin{eqnarray}\label{}
\int dx\,\rho(x) &=&\frac{2\pi}{\sqrt{\lambda}}(\Delta+2M-L),
\\
\int dx\,\frac{\rho(x)}{x}&=&2\pi m,
\\ \label{lnorm}
\int dx\,\frac{\rho(x)}{x^2}
&=&\frac{2\pi}{\sqrt{\lambda}}(\Delta-L).
\end{eqnarray}

The equations (\ref{sbethe})--(\ref{lnorm}) are very similar to the
classical Bethe equations for the spin chain. The change of
variables $x\rightarrow 4\pi L x/\sqrt{\lambda}$ explicitly relates
the two sets of equations:
\begin{eqnarray}\label{}
2\pint dy\,\frac{\rho(y)}{x-y}&=&\frac{x}{x^2-\frac{\lambda }{16\pi
^2L ^2}}\,\,\frac{\Delta}{L} +2\pi n_k, \qquad x\in C_k,
\\
\int dx\,\rho (x) &=&\frac{M}{L}+\frac{\Delta-L}{2L }\,,
\\
\int dx\,\frac{\rho (x)}{x}&=&2\pi m,
\\
\Delta -L&=&\frac{\lambda }{8\pi ^2L}\int  dx\,\frac{\rho
(x)}{x^2}\,.
\end{eqnarray}
The one-loop classical Bethe ansatz (\ref{bte}), (\ref{nom}),
(\ref{mom}) and (\ref{adim}) is recovered in the limit
$\lambda/L^2\rightarrow 0$.

\section{Discussion}

The main result of the long derivation in sec.~\ref{adssec} is an
integral equation that parameterizes classical solutions of the
sigma-model \cite{Kazakov:2004qf}:
\begin{equation}\label{1}
2\pint dy\,\frac{\rho(y)}{x-y}=\frac{x}{x^2-\frac{\lambda }{16\pi
^2L ^2}}\,\,\frac{\Delta}{L} +2\pi n_k.
\end{equation}
This equation reduces at $\lambda/L^2\rightarrow 0$ to the classical
Bethe equation for the spin chain:
\begin{equation}\label{2}
2\pint dy\,\frac{\rho(y)}{x-y}=\frac{1}{x}+2\pi n_k.
\end{equation}
That equation in turn is an approximation to the exact quantum Bethe
equations
\begin{equation}\label{3}
\left(u_j+i/2\over u_j-i/2\right)^L= \prod_{k\ne j}
 {u_j-u_k+i\over
u_j-u_k-i}\,.
\end{equation}
 In view of the analogy between (\ref{1}) and (\ref{2}), it is
natural to ask if (\ref{1}) is a scaling limit of some discrete
Bethe equations which describe the full quantum spectrum of the AdS
string. Potentially useful analogy in this respect is the $SU(2)$
chiral field (sigma-model on $S^3$) which classically is a subsector
of the full \ads\ sigma-model. The quantum chiral field can be
fermionized and solved by Bethe ansatz \cite{Polyakov:1983tt}.
Perhaps the sigma-model in \ads\ can also be fermi/bosonized and
solved by similar techniques. Another possibility is that the
semiclassical approximation and symmetries uniquely fix quantum
Bethe equations. This does not look completely inconceivable, since
for many integrable systems understanding the semiclassical limit is
sufficient for quantization. The quantum string Bethe ansatz should
constitute a set of algebraic equations for momenta of elementary
excitations on the world-sheet and should reduce to (\ref{1}) in the
semiclassical limit. A particular discretization of (\ref{1}) was
proposed in \cite{Arutyunov:2004vx} and passed a number of highly
non-trivial checks. The equations of \cite{Arutyunov:2004vx} are
still valid only at strong coupling, but they do have a spin-chain
interpretation \cite{Beisert:2004jw}. This may suggest a way to
compare the string Bethe ansatz with the Bethe ansatz for
perturbative SYM theory.

\section*{Acknowledgements}

I would like to thank V.~Kazakov, A.~Marshakov and J.~Minahan for
fruitful collaboration and A.~Niemi for explaining me the Berry's
phase.




\end{document}